\documentclass[preprint]{revtex4}

\usepackage{graphicx}

\newcommand{\E}{{\mathcal{E}}}
\newcommand{\s}{\sigma}

\renewcommand{\a}{\alpha}

\newcommand{\q}{{\sf{q}}}

\newcommand{\be}{\begin{equation}}
\newcommand{\ee}{\end{equation}}
\newcommand{\bea}{\begin{eqnarray}}
\newcommand{\eea}{\end{eqnarray}}
\newcommand{\ba}{\begin{array}}
\newcommand{\ea}{\end{array}}
\def\J#1#2#3#4{{#1} {\bf #2}, #3 (#4)}
\def\PRD{Phys. Rev. D}
\def\PR{Phys. Rev.}
\def\PRL{Phys. Rev. Lett.}
\def\PTP{Prog. Theor. Phys.}

\def\APL{Ann. Phys. (Leipzig)}

\def\JMP{J. Math. Phys.}

\def\TMP{Theor. Math. Phys.}

\def\CQG{Class. Quantum Grav.}

\def\PLB{Phys. Lett. B}
\def\PTP{Prog. Theor. Phys.}
\def\PTEP{Prog. Theor. Exper. Phys.}
\def\PRSLA{Proc. R. Soc. Lond. A}

\def\NP{Nucl. Phys.}

\begin{document}
\draft
\title{Stationary black diholes}

\author{V.~S.~Manko$^\dag$, R. I. Rabad\'an$^\dag$ and J. D. Sanabria-G\'omez$^\ddag$}
\address{$^\dag$Departamento de F\'\i sica, Centro de Investigaci\'on y de
Estudios Avanzados del IPN, A.P. 14-740, 07000 M\'exico D.F.,
Mexico\\$^\ddag$Escuela de F\'isica, Universidad Industrial de
Santander, A.A. 678, Bucaramanga, Colombia}

\begin{abstract}
In this paper we present and analyze the simplest physically
meaningful model for stationary black diholes -- a binary
configuration of counter-rotating Kerr-Newman black holes endowed
with opposite electric charges -- elaborated in a physical
parametrization on the basis of one of the Ernst-Manko-Ruiz
equatorially antisymmetric solutions of the Einstein-Maxwell
equations. The model saturates the Gabach-Clement inequality for
interacting black holes with struts, and in the absence of
rotation it reduces to the Emparan-Teo electric dihole solution.
The physical characteristics of each dihole constituent satisfy
identically the well-known Smarr's mass formula.
\end{abstract}


\maketitle

%\twocolumn

\section{Introduction}

In the paper \cite{ETe} Emparan and Teo constructed and analyzed
for the first time an exact electrostatic solution of the
Einstein-Maxwell equations describing a non-extremal dihole -- a
configuration consisting of two non-extremal Reissner-Nordstr\"om
black holes \cite{Rei,Nor} endowed with equal masses and opposite
charges of the same magnitude. The black-hole constituents in the
Emparan-Teo (ET) solution are prevented from falling onto each
other by a massless strut (its rough Newtonian analog is a thin
rod whose mass can be neglected) the pressure inside of which
permits one to get information about the interaction force between
the constituents. Various thermodynamical properties of the
interacting static black holes were also studied in \cite{ETe},
which became possible thanks to a known nice physical property of
the spacetimes with struts -- the struts do not contribute into
the total gravitational energy of the system. At the end of their
paper, Emparan and Teo mentioned that an extension of their
results to the case of rotation might be an interesting work to be
done in the future, and the problem of obtaining such an extension
seems to have been attacked in a recent paper of Cabrera-Munguia
et al \cite{CLL} who constructed a specific 4-parameter exact
solution for two oppositely charged counter-rotating Kerr-Newman
(KN) black holes \cite{NCC}. Though technically the paper
\cite{CLL} is correct, the solution itself, in our opinion,
exhibits some unphysical features because of the presence in it of
non-vanishing magnetic charges created by rotation of electric
charges, which contradicts the known cases of a single KN solution
and of the Bret\'on-Manko (BM) solution \cite{BMa,MRR} for a pair
of identical counter-rotating KN black holes where the electric
charges generate a {\it dipole} magnetic field without magnetic
monopoles. As a consequence, the usual Smarr mass formula
\cite{Sma} (not taking into account the contribution of magnetic
charges) does not hold for the black-hole constituents comprising
that binary configuration and, moreover, the expression of the
important geometric quantity $\s$ obtained in \cite{CLL} depends
implicitly on the root of a cubic algebraic equation, being
therefore more complicated than for instance the analogous
quantity of the physically parametrized BM solution.

The present paper aims at working out a unique 4-parameter model
for stationary diholes with antiparallel rotation of its
constituents, in which would be absent not only the total but also
individual magnetic charges. To accomplish this task, we shall
make use of a 5-parameter asymptotically flat specialization of
the Ernst-Manko-Ruiz (EMR) equatorially antisymmetric solution
\cite{EMR1,SRo} possessing arbitrary parameters of electric and
magnetic dipole moments, which will enable us to eliminate the
individual magnetic charges of the constituents by choosing
appropriately the values of the latter dipole parameters. This
will secure the validity of the standard Smarr formula for each
black-hole constituent and, in turn, will enable us to find a
remarkably simple expression for $\s$ in terms of the Komar
quantities \cite{Kom} which is a key point for elaborating the
physical parametrization of the whole model. After having reached
our main objective, we will prove that, similar to the BM model of
equally charged counter-rotating black-hole constituents, the
configuration obtained for KN black holes with opposite electric
charges verifies (and actually saturates) the inequality for
interacting black holes with struts recently derived by Gabach
Clement \cite{Gab}.

\section{The 5-parameter asymptotically flat EMR solution\\
in $\s$-representation}

A key result of the paper \cite{CLL} is its expressions (14) for
the axis data $e(z)$ and $f(z)$ allowing one to construct in a
straightforward manner the corresponding entire metric with the
aid of the general formulas of the extended $N$-soliton solution
\cite{RMM} (we also refer the interested reader to Appendix of
Ref.~\cite{SRo} for a complete set of algebraic relations involved
in the construction procedure of the $N=2$ case). However, an
explanation the authors of \cite{CLL} give to the origin of those
expressions -- the solution (11) of a complicated system of
algebraic equations for certain metric functions -- raises in turn
the question of how these formulas~(11) were obtained, and below
we will give a simple derivation of their data~(14) with one
additional arbitrary real parameter representing a magnetic dipole
moment.

As a starting point of the derivation procedure we take the
following axis data obtained in the paper~\cite{EMR1} for an
equatorially antisymmetric spacetime \cite{EMR2} with both
electric and magnetic dipole moments:
\be e(z)=\frac{z^2-b_1z+b_2}{z^2+b_1z+b_2}, \quad
f(z)=\frac{c_2}{z^2+b_1z+b_2}, \label{ad6} \ee
where $b_1$, $b_2$ and $c_2$ are arbitrary complex constants.
Mention that this data rewritten in an equivalent representation
was used in the paper \cite{EMR1} for constructing the
corresponding Ernst potentials \cite{Ern} $\E$ and $\Phi$ of a
6-parameter EMR solution. If nevertheless one opts to work
directly with the above (\ref{ad6}), then the asymptotic flatness
of the solution implies immediately that $b_1$ is a real constant
related to the total mass $2M$ of the binary configuration as
$b_1=2M$. Choosing then the constant $c_2$ in the form
$c_2=2(q+ib)$, the real parameters $q$ and $b$ being associated,
respectively, with the electric and magnetic dipole moments, and
also formally setting $b_2=c-i\delta$, we arrive at the
5-parameter axis data
\be e(z)=\frac{e_-}{e_+}, \quad f(z)=\frac{2(q+ib)}{e_+}, \quad
e_\mp=z^2\mp 2Mz+c-i\delta, \label{ad5} \ee
in which the real constants $c$ and $\delta$ should yet be related
to some physical or geometrical characteristics.

Recall now that the extended multi-soliton solutions involve the
constants $\a_n$ which satisfy the algebraic equation \cite{Sib}
\be e(z)+\bar e(z)+2f(z)\bar f(z)=0 \label{sib} \ee
(the bar over a symbol means complex conjugation), and in the
equatorially antisymmetric case these can be chosen in the form
\be \a_1=-\a_4=\frac{1}{2}R+\s, \quad \a_2=-\a_3=\frac{1}{2}R-\s,
\label{alf} \ee
where $R$ is a real constant representing the coordinate
separation of the sources, and the parameter $\s$ can take
non-negative real or pure imaginary values (see Fig.~1). Then
instead of the constants $c$ and $\delta$ from the axis data
(\ref{ad5}) one is able to introduce the new parameters $R$ and
$\s$ by equating coefficients at the same powers of $z$ on the two
sides of the equation
\be \frac{e_-}{e_+}+\frac{\bar e_-}{\bar
e_+}+\frac{4(q^2+b^2)}{e_+\bar e_+}=
\frac{2\prod_{n=1}^4(z-\a_n)}{e_+\bar e_+}, \label{eq} \ee
thus yielding
\be c=2M^2-\frac{1}{4}R^2-\s^2, \quad
\delta=\sqrt{(R^2-4M^2)(M^2-\s^2)-4(q^2+b^2)}. \label{cd} \ee

Accounting for (\ref{cd}), the 5-parameter axis data (\ref{ad5})
finally assumes the form
\be e(z)=\frac{e_-}{e_+}, \quad f(z)=\frac{2(q+ib)}{e_+}, \quad
e_\mp=z^2\mp 2Mz+2M^2-\frac{1}{4}R^2-\s^2-i\delta, \label{ad5n}
\ee
and, by setting $b=0$ in (\ref{cd}) and (\ref{ad5n}), one recovers
the axis data (14) of \cite{CLL} (and consequently the quantities
$\beta_{1,2}$ and $f_{1,2}$ in (11) of \cite{CLL} via the simple
fraction decomposition of $e(z)$ and $f(z)$). It is worth noting
that this procedure of changing parameters in the axis data was
already described in application to the case of identical
counter-rotating uncharged black holes \cite{MRRS} and, moreover,
has been recently used for obtaining a physical parametrization of
the BM solution \cite{MRR}. Furthermore, by virtue of the
equatorial antisymmetry, the axis condition for the solution
defined by the axis data (\ref{ad5n}) is satisfied automatically,
and therefore there is no need to solve any additional algebraic
equations for the metric functions.

As it is straightforward to elaborate by purely algebraic
computing the explicit form of the Ernst potentials defined by the
axis data (\ref{ad5n}), as well as the form of the corresponding
metric functions $f$, $\gamma$ and $\omega$ entering the
stationary axisymmetric line element
\be d s^2=f^{-1}[e^{2\gamma}(d\rho^2+d z^2)+\rho^2d\varphi^2]-f(d
t-\omega d\varphi)^2, \label{pap} \ee
below we will restrict ourselves to only writing out the final
expressions which reproduce and generalize the analogous formulas
of the paper \cite{CLL}. Then for $\E$ and $\Phi$ we have
\bea \E&=&\frac{A-B}{A+B}, \quad \Phi=\frac{C}{A+B}, \nonumber\\
A&=&R^2[M^2(R^2-4\s^2)-4(q^2+b^2)](R_+-R_-)(r_+-r_-)+
4\s^2[M^2(R^2-4\s^2) \nonumber\\
&&+4(q^2+b^2)](R_+-r_+)(R_--r_-)+2R\s(R^2-4\s^2)[R\s(R_+r_-+R_-r_+)
\nonumber\\ &&+i\delta(R_+r_--R_-r_+)], \nonumber\\
B&=&2MR\s(R^2-4\s^2)[R\s(R_++R_-+r_++r_-)-(2M^2-i\delta)(R_+-R_--r_++r_-)],
\nonumber\\
C&=&4(q+ib)R\s[(R+2\s)(R\s-2M^2-i\delta)(r_+-R_-)
+(R-2\s)(R\s+2M^2+i\delta) \nonumber\\ &&\times(r_--R_+)],
\label{EF} \eea
where
\be R_\pm=\sqrt{\rho^2+(z+{\textstyle\frac{1}{2}}R\pm\sigma)^2},
\quad r_\pm=\sqrt{\rho^2+(z-{\textstyle\frac{1}{2}}R\pm\sigma)^2},
\label{Rpm} \ee
while the metric functions are given by the expressions
\bea f&=&\frac{A\bar A-B\bar B+C\bar C}{(A+B)(\bar A+\bar B)},
\quad e^{2\gamma}=\frac{A\bar A-B\bar B+C\bar C}
{16R^4\s^4(R^2-4\s^2)^2R_+R_-r_+r_-}, \nonumber\\
\omega&=&-\frac{{\rm Im}[2G(\bar A+\bar B)+C\bar I]}
{A\bar A-B\bar B+C\bar C}, \nonumber \\
G&=&-zB+R\s\{2R[M^2(R^2-4\s^2)-2(q^2+b^2)](R_-r_--R_+r_+) \nonumber\\
&&+4\s[M^2(R^2-4\s^2)+2(q^2+b^2)](r_+r_--R_+R_-)
\nonumber\\
&&+M(R+2\s)[(R-2\s)^2(R\s+2M^2-i\delta)-8(q^2+b^2)](R_--r_+)
\nonumber\\
&&+M(R-2\s)[(R+2\s)^2(R\s-2M^2+i\delta)+8(q^2+b^2)](R_+-r_-)\},
\nonumber\\
I&=&-zC+4M(q+ib)[R^2(2M^2-2\s^2+i\delta)(R_+r_++R_-r_-)
\nonumber\\
&&+2\s^2(R^2-4M^2-2i\delta)(R_+R_-+r_+r_-)]-2(q+ib)(R^2-4\s^2)
\nonumber\\ &&\times\{2M[(R\s+2M^2+i\delta)R_+r_-
-(R\s-2M^2-i\delta)R_-r_+] +R\s \nonumber\\
&&\times[(R\s+6M^2+i\delta)(R_++r_-)+ (R\s-6M^2-i\delta)(R_-+r_+)
+8MR\s]\}. \label{mf} \eea

The $t$ and $\varphi$ components of the electromagnetic
4-potential are defined by the formulas
\be A_t=-{\rm Re}\left(\frac{C}{A+B}\right), \quad A_\varphi={\rm
Im}\left(\frac{I}{A+B}\right), \label{tph} \ee
and these complete the general mathematical description of the
5-parameter EMR solution in $\s$-parametrization.

At this point, several remarks on the formulas
(\ref{EF})-(\ref{tph}) might be appropriate. First, the above
representation of the 5-parameter EMR solution is fully equivalent
to the known description of that solution worked out in the papers
\cite{EMR1,SRo} (with the NUT parameter $\nu$ set equal to zero).
Second, it is highly important to underline that the arbitrary
parameter $\s$ of the solution is not restricted exclusively to
real values (contrary to what was assumed in \cite{CLL}) but can
also take pure imaginary values determining the hyperextreme part
of the solution. The significance of this point will be fully
understandable later on when we express $\s$ in terms of the Komar
quantities. Third, the $\s$- representation of the EMR solution
should be only considered as an intermediate parametrization that
could be suitable for elaborating the final physical
representation in which $\s$ must be replaced by a rotation
parameter.

To gain a better insight into the structure of the EMR solution,
let us consider its first four Beig-Simon multipole moments
\cite{BSi,SBe,Sim} which can be found with the aid of the
Hoenselaers-Perj\'es procedure \cite{HPe,SAp}:
\bea M_0&=&2M, \quad M_1=0, \quad
M_2=\frac{1}{2}M(R^2-8M^2+4\s^2), \quad M_3=0, \nonumber \\
J_0&=&J_1=0, \quad J_2=2M\delta, \quad J_3=0, \nonumber \\
Q_0&=&0, \quad Q_1=2q, \quad Q_2=0, \quad
Q_3=\frac{1}{2}q(R^2-8M^2+4\s^2) -2b\delta, \nonumber \\ B_0&=&0,
\quad B_1=2b, \quad B_2=0, \quad B_3=\frac{1}{2}b(R^2-8M^2+4\s^2)
+2q\delta \label{mm} \eea
($M_i$, $J_i$, $Q_i$ and $B_i$ define, respectively, the mass,
angular momentum, electric and magnetic multipole moments), whence
it follows the asymptotic flatness of the solution ($J_0=0$), the
total mass $M_0$ of the configuration being equal to $2M$, and
total angular momentum $J_1$ being zero due to counterrotation. In
the absence of net charges the parameters $q$ and $b$ define the
electric and magnetic dipole moments, respectively, which means
that the two sources in the EMR solution are endowed with opposite
electric and magnetic charges.

It is clear from the above form of the multipole moments that the
special $b=0$ case of the EMR solution considered in \cite{CLL} is
characterized by zero total magnetic dipole moment, and this fact
explains the intrinsic presence of magnetic monopoles in that
particular solution. Indeed, the magnetic dipole moment $2b$ of
the 5-parameter EMR solution is a result of the following two
non-zero contributions -- one coming from the rotating electric
charges and the other originated by the opposite magnetic charges.
The electric contribution is twice the magnetic dipole moment
created by one rotating electric charge, so by demanding $b=0$,
Cabrera-Munguia et al introduced in \cite{CLL} a specific
non-vanishing magnetic dipole moment due to magnetic charges,
antiparallel to that created by electric charges. It would be
plausible to suppose that those authors probably confused the case
of counter-rotating opposite charges with the BM configuration in
which the counter-rotating charges have the same signs and hence
the total magnetic and electric dipole moments are both equal to
zero intrinsically. Therefore, a physically meaningful dihole
solution arising from the 5-parameter EMR configuration must have
zero individual magnetic charges and, at the same time, a non-zero
magnetic dipole moment generated by counterrotation of opposite
electric charges.

The individual magnetic charges in the 5-parameter EMR solution
can be eliminated by means of the condition \cite{Tom,MMR}
\be A_t(\rho=0,z=\a_1)-A_t(\rho=0,z=\a_2)=0, \label{cond} \ee
which can be easily solved for $b$. Then from (\ref{EF}),
(\ref{Rpm}) and (\ref{tph}) we get
\be
b^2=\frac{4q^2[(R^2-4M^2)(M^2-\s^2)-4q^2]}{(R^2-4M^2)^2+16q^2},
\label{b} \ee
and this condition, together with the formulas
(\ref{EF})-(\ref{tph}) with real $\s$, provide one with a
$\s$-representation of the physically meaningful 4-parameter model
for a stationary black dihole whose constituents are
counter-rotating. It can be verified by a direct calculation that
each black-hole constituent of such a model verifies the
well-known Smarr mass formula identically.

\section{The 4-parameter dihole solution in physical\\ parametrization}

As was already mentioned, the $\s$-representation of the solution
is only an intermediate step on the way of obtaining the physical
parametrization in terms of the Komar quantities. Once the
$\s$-representation is known, our further actions are the
following: we must first try to express the parameter $q$ in terms
of the individual Komar charge $Q$ of any of the dihole
constituents, thus rewriting the solution in the parameters $M$,
$R$, $Q$ and $\s$, and then find the form of $\s$ in terms of $M$,
$R$, $Q$ and $J$, $J$ being the individual Komar angular momentum,
from Smarr's mass formula, by considering the latter an algebraic
equation for $\s$. Mention here that although the coordinate
distance $R$ is not an invariantly defined quantity, its
introduction instead of the proper distance integral
$\int\sqrt{f^{-1}\exp(2\gamma)}|_{\rho=0}dz$ is justified by the
possibility to obtain simple analytical formulas very suitable for
carrying out the physical analysis.

The mass formula for black holes discovered by Smarr \cite{Sma}
relates the mass $M$, angular momentum $J$ and charge $Q$ of a
black hole to several quantities evaluated on the horizon: the
surface gravity $\kappa$, horizon's area $S$ and angular velocity
$\Omega^H$, and the electric potential $\Phi^H$. The formula reads
\be M=\frac{1}{4\pi}\kappa S+2J\Omega^H+Q\Phi^H
=\s+2J\Omega^H+Q\Phi^H, \label{Sma} \ee
the Komar quantities $M$, $J$ and $Q$ being defined by the
integrals \cite{Tom}
\bea M&=&-\frac{1}{8\pi}\int_{H}
\omega\Omega_{,z} d\varphi d z, \label{kq1}\\
J&=&\frac{1}{8\pi}\int_{H}\omega
\left[-1-{\textstyle\frac12}\omega\Omega_{,z}+\tilde
A_\varphi A'_{\varphi,z}+(A_\varphi A'_\varphi)_{,z}\right] d\varphi d z, \label{kq2}\\
Q&=&\frac{1}{4\pi}\int_{H} \omega A'_{\varphi,z} d\varphi d z,
\label{kq3} \eea
with $\Omega={\rm Im}(\E)$, $A'_\varphi={\rm Im}(\Phi)$, $\tilde
A_\varphi=A_\varphi+\omega A_t$ (note that the metric functions
$\omega$ and $\gamma$, as well as the potential $\tilde
A_\varphi$, take constant values on the horizon), while the form
of the constants $\kappa$, $S$, $\Omega^H$ and $\Phi^H$ is given
by the formulas \cite{Tom,Car}
\be \kappa=\sqrt{-\omega^{-2}e^{-2\gamma}}, \quad
S=4\pi\sigma\sqrt{-\omega^{2}e^{2\gamma}}, \quad
\Omega^H=\omega^{-1}, \quad \Phi^H=-A_t-\Omega^H A_\varphi.
\label{kap} \ee

For our dihole solution, the calculation of the individual charge
$Q$ of the upper black-hole constituent, whose horizon is
represented by the null hypersurface $\rho=0$,
${\textstyle\frac12}R-\s\le z\le {\textstyle\frac12}R+\s$, leads
to a cubic equation for $q$ which has to be solved in order to
pass from the latter $q$ to the Komar $Q$ in the formulas
determining the solution. It is remarkable, however, that the need
to solve a cubic equation can be circumvented by an appropriate
change of the parameter $q$. Thus, by introducing a new parameter
$\q$ via the relation
\be \q^2=q^2+b^2, \label{qnew} \ee
which has some analogy with a duality rotation of the
electromagnetic Ernst potential, we find from (\ref{b}) and
(\ref{qnew}) the form of $q$ and $b$ in terms of $\q$:
\bea q&=&\q(R^2-4M^2)/\tau, \quad b=2\q
\delta'/\tau, \nonumber\\
\delta'&=&\sqrt{(R^2-4M^2)(R^2-\s^2)-4\q^2}, \quad
\tau=\sqrt{(R^2-4M^2)(R^2-4\s^2)-16\q^2}, \label{qb} \eea
and this redefinition of the parameter $q$ permits us to obtain
from (\ref{kq3}) a simple expression for the Komar charge $Q$ in
terms of $\q$:
\be Q=-2\q(R+2M)/\tau, \label{QK} \ee
whence we readily get the inverse dependence of $\q$ on $Q$:
\be
\q=-\frac{Q\sqrt{(R^2-4M^2)(R^2-4\s^2)}}{2\sqrt{(R+2M)^2+4Q^2}}.
\label{qQ} \ee

The above formula for $\q$ permits us, by rewriting the dihole
solution in terms of the parameters $M$, $R$, $Q$ and $\s$, to
obtain the quantities $\Omega^H$ and $\Phi^H$ that we need for
finding $\s$:
\bea &&\Omega^H=\frac{\sqrt{(R-2M)[(R+2M)^2+4Q^2]
[(R+2M)(M^2-\s^2)-Q^2(R-2M)]}}{(R+2M)[2M(R+2M)(M+\s)-Q^2(R-4M-2\s)]},
\nonumber\\ &&\Phi^H=\frac{Q(R-2M)[(R+2M)(M+\s)+2Q^2]
}{(R+2M)[2M(R+2M)(M+\s)-Q^2(R-4M-2\s)]}. \label{OF} \eea

Finally, after the substitution of (\ref{OF}) into the Smarr
formula (\ref{Sma}) in which we can put $J=Ma$, $a$ being the
angular momentum per unit mass of the upper black hole, we obtain
by a simple calculation the desired expression for $\s$ in terms
of the physical quantities $M$, $a$, $Q$ and $R$:
\be \s=\sqrt{M^2-\left(\frac{M^2a^2[(R+2M)^2+4Q^2]}
{[M(R+2M)+Q^2]^2}+Q^2\right)\frac{R-2M}{R+2M}}. \label{sig} \ee

This formula for $\s$ is the {\it central} result of our paper.
Now the dihole solution can be rewritten in the physical
parameters, its Ernst potentials $\E$ and $\Phi$ assuming the form
\bea \E&=&\frac{A-B}{A+B}, \quad \Phi=\frac{C}{A+B}, \nonumber\\
A&=&R^2(M^2-Q^2\nu)(R_+-R_-)(r_+-r_-)+
4\s^2(M^2+Q^2\nu)(R_+-r_+)(R_--r_-) \nonumber\\
&&+2R\s[R\s(R_+r_-+R_-r_+)+iMa\mu(R_+r_--R_-r_+)], \nonumber\\
B&=&2MR\s[R\s(R_++R_-+r_++r_-)-(2M^2-iMa\mu)(R_+-R_--r_++r_-)],
\nonumber\\
C&=&2C_0R\s[(R+2\s)(R\s-2M^2-iMa\mu)(r_+-R_-) +(R-2\s) \nonumber\\
&&\times(R\s+2M^2+iMa\mu)(r_--R_+)], \label{EFn} \eea
where the dimensionless quantities $\mu$, $\nu$ and $C_0$ are
defined as
\be \mu=\frac{R^2-4M^2}{M(R+2M)+Q^2}, \quad
\nu=\frac{R^2-4M^2}{(R+2M)^2+4Q^2}, \quad
C_0=-\frac{Q(R^2-4M^2+2iMa\mu)}{(R+2M)(R^2-4\s^2)}, \label{mnc}
\ee
and the final form of the metric coefficients $f$, $\gamma$ and
$\omega$ is the following:
\bea f&=&\frac{A\bar A-B\bar B+C\bar C}{(A+B)(\bar A+\bar B)},
\quad e^{2\gamma}=\frac{A\bar A-B\bar B+C\bar C}
{16R^4\s^4R_+R_-r_+r_-}, \quad
\omega=-\frac{{\rm Im}[2G(\bar A+\bar B)+C\bar I]}
{A\bar A-B\bar B+C\bar C}, \nonumber \\
G&=&-zB+R\s\{R(2M^2-Q^2\nu)(R_-r_--R_+r_+)
+2\s(2M^2+Q^2\nu)(r_+r_--R_+R_-)
\nonumber\\
&&+M[(R+2\s)(R\s-2M^2+iMa\mu)+2(R-2\s)Q^2\nu](R_+-r_-)
\nonumber\\
&&+M[(R-2\s)(R\s+2M^2-iMa\mu)-2(R+2\s)Q^2\nu](R_--r_+)\},
\nonumber\\
I&=&-zC+2C_0M[R^2(2M^2-2\s^2+iMa\mu)(R_+r_++R_-r_-)
\nonumber\\
&&+2\s^2(R^2-4M^2-2iMa\mu)(R_+R_-+r_+r_-)]-C_0(R^2-4\s^2) \nonumber\\
&&\times\{2M[R\s(R_+r_--R_-r_+)+(2M^2+iMa\mu)(R_+r_-+R_-r_+)] +R\s[R\s \nonumber\\
&&\times(R_++R_-+r_++r_-)+(6M^2+iMa\mu)(R_+-R_--r_++r_-)
+8MR\s]\}. \label{mfn} \eea

It should be noted that $\s$ in the above (\ref{EFn})-(\ref{mfn})
is no longer an independent parameter, having conceded that role
to the constant $a$. From (\ref{sig}) it follows that $\s$,
depending on interrelations between the parameters $M$, $a$, $Q$
and $R$, can automatically take on (non-negative) real or pure
imaginary values, thus describing not only the binary
configurations of black holes but also of hyperextreme objects.
That is why $\s$'s taking pure imaginary values (along with the
real ones) in the EMR solution (\ref{EF}) is highly important for
the mathematical equivalence of the parameter sets ($M$, $Q$, $R$,
$\s$) and ($M$, $Q$, $R$, $a$), and consequently for the
correctness of the entire reparametrization procedure. However,
since our primary interest lies in the black-hole sector
($\s\ge0$) of the dihole solution, we may always restrict
ourselves to those values of the physical parameters $M$, $Q$, $R$
and $a$ that preserve the reality of $\s$.

\section{The limits and physical properties of dihole solution}

The main limits of the dihole solution can be well seen from the
formula (\ref{sig}) for $\s$. Thus, in the absence of rotation
($a=0$) the solution reduces to the Emparan-Teo electrostatic
non-extreme dihole spacetime \cite{ETe} whose physical form was
found in the paper \cite{CGM}. In the pure vacuum limit ($Q=0$)
the solution represents a vacuum specialization of the BM
equatorially antisymmetric binary configuration whose physical
parametrization was elaborated in the paper \cite{MRRS} on the
basis of Varzugin's expression for the quantity $\s$ \cite{Var}.
When $R\to\infty$ (no interaction between the dihole
constituents), one gets from (\ref{sig}) $\s=(M^2-a^2-Q^2)^{1/2}$,
which is characteristic of a single KN black hole.

By construction, the upper KN constituent has mass $M$, angular
momentum $Ma$ and charge $Q$, whereas the analogous
characteristics of the lower constituent are $M$, $-Ma$ and $-Q$,
respectively. The strut separating the two constituents provides
us with the information about the interaction force \cite{Isr},
the latter being defined by the expression
\bea {\mathcal
F}=\frac14(e^{-\gamma_0}-1)&=&\frac{M^2(R+2M)^2+Q^2R^2}
{(R+2M)^2(R^2-4M^2)} \nonumber\\
&=&\frac{1}{R^2-4M^2}\left(M^2+Q^2-
\frac{4MQ^2(R+M)}{(R+2M)^2}\right), \label{F} \eea
where $\gamma_0$ is the value of the metric function $\gamma$ on
the strut, and one can see that ${\mathcal F}$ cannot take zero
value at any finite separation $R$ of the constituents, so that
the strut is irremovable generically. It is worth mentioning that
formula (\ref{F}) differs from the analogous expression obtained
in \cite{CLL}, as our ${\mathcal F}$ does not contain a
contribution due to magnetic charges. Note also that ${\cal F}$
admits a geometrical interpretation in terms of the conical
deficit quantity $\delta$ via the formula ${\cal
F}=-\delta/(8\pi)$, and that the absence of the angular momentum
$J$ in the formula (\ref{F}) does not mean at all that the
spin-spin interaction is not present in our dihole model since, as
has already been rightly observed in the paper \cite{CHR}, the
angular momenta contribution (attractive) will enter explicitly
into the above formula after using in the latter of the proper
distance integral instead of the coordinate distance $R$.

Turning now to the thermodynamical quantities $\kappa$, $S$,
$\Omega^H$ and $\Phi^H$ entering the Smarr mass formula
(\ref{Sma}), it may be observed that these must be calculated only
for the upper black-hole constituent because the analogous set for
the lower constituent is just $\kappa$, $S$, $-\Omega^H$ and
$-\Phi^H$. Then for the upper constituent we get
\bea \kappa&=&\frac{R\s[(R+2M)^2+4Q^2]}
{(R+2M)^2[2(M+\s)(MR+2M^2+Q^2)-Q^2(R-2M)]}, \label{kS1}\\
S&=&4\pi\left(1+\frac{2M}{R}\right) \left(2M(M+\s)-
\frac{Q^2(R-2M)(R-2\s)}{(R+2M)^2+4Q^2}\right) \nonumber \\
&=&\frac{4\pi}{R(R+2\s)}\left((R+2M)^2(M+\s)^2
+\frac{M^2a^2(R^2-4M^2)^2}{(MR+2M^2+Q^2)^2}\right), \label{kS2}\\
\Omega^H&=&\frac{Ma[2(M-\s)(MR+2M^2+Q^2)-Q^2(R-2M)]}
{(4M^2a^2+Q^4)(MR+2M^2+Q^2)}, \label{kS3}\\
\Phi^H&=&\frac{Q[Q^2(M-\s)(MR+2M^2+Q^2)+2M^2a^2(R-2M)]}
{(4M^2a^2+Q^4)(MR+2M^2+Q^2)},  \label{kS4} \eea
where $S$ is given in two different forms suitable for recovering
the known limiting cases. The substitution of
(\ref{kS1})-(\ref{kS4}) into (\ref{Sma}) shows that the above
expressions satisfy identically Smarr's formula for black holes.

It has been recently shown \cite{MRR} that the equally charged
black-hole constituents of the BM configuration saturate the
Gabach-Clement inequality for black holes with struts \cite{Gab}
which reads
\be \sqrt{1+4{\mathcal F}}\ge\frac{\sqrt{(8\pi J)^2+(4\pi
Q^2)^2}}{S}. \label{GC} \ee
In this respect it would be interesting to clarify whether the
oppositely charged constituents of our dihole model saturate the
inequality (\ref{GC}) too. The saturation means that the extremal
dihole constituents must satisfy (\ref{GC}) with the equality
sign. The extremality condition $\s=0$ yields from (\ref{sig}) the
value of $a$ at which the black-hole degeneration occurs:
\be a^2=\frac{(MR+2M^2+Q^2)^2[M^2(R+2M)-Q^2(R-2M)]}
{M^2(R-2M)[(R+2M)^2+4Q^2]}, \label{ext} \ee
and by substituting this $a$ into (\ref{kS2}) and (\ref{GC}) we
get ($J=Ma$)
\bea &&S=\frac{4\pi(R+2M)^2[2M^2(R+2M)-Q^2(R-4M)]}
{R[(R+2M)^2+4Q^2]}, \nonumber\\ &&(8\pi J)^2+(4\pi Q^2)^2
=\frac{16\pi^2(R+2M)[2M^2(R+2M)-Q^2(R-4M)]^2}
{(R-2M)[(R+2M)^2+4Q^2]}. \label{SI}   \eea

Taking into account that ${\mathcal F}$ does not depend explicitly
on $a$, it is easy to check that (\ref{F}) and (\ref{SI}) verify
the equality in (\ref{GC}) identically. Therefore, independently
of whether the KN black holes in a binary system have equal or
opposite charges, the interaction force between them is governed
by the Gabach-Clement inequality. Mention here one more common
feature shared by the BM and dihole configurations -- the extreme
limit is achieved in both of them at a larger absolute value of
$a$ (for some given $M$ and $Q$) than in the case of a single KN
black hole whose extremality condition is simply $a^2=M^2-Q^2$.

Mention that in the paper \cite{CLL} the Ernst potentials defining
the extreme limit of the 4-parameter solution are given with
errors. Therefore, we find it useful to give below the expressions
for these potentials and corresponding metric functions of the
entire 5-parameter EMR solution in the extreme limit $\s\to 0$:
\bea \E&=&\frac{A-B}{A+B}, \quad \Phi=\frac{C}{A+B}, \nonumber \\
f&=&\frac{N}{D}, \quad e^{2\gamma}=\frac{N} {\a^8(x^2-y^2)^4},
\quad \omega=-\frac{4\a^2\delta y(x^2-1)(1-y^2)W} {N}, \nonumber\\
A&=&M^2\a^2(x^4-1)+\a^2(\a^2-M^2)(x^2-y^2)^2+(q^2+b^2)(1-y^4)
\nonumber\\ &&+2i\a^2\delta(x^2-2x^2y^2+y^2), \nonumber\\ B&=&2M\a
x[\a^2(x^2-y^2)-(M^2-i\delta)(1-y^2)], \nonumber\\ C&=&-2(q+ib)y
[\a^2(x^2-y^2)-(M^2+i\delta)(1-y^2)], \nonumber\\
I&=&-\a xyC-2(q+ib)(M+\a x)(1-y^2) [(M+\a x)^2+(M^2-\a^2)y^2
+i\delta(1+y^2)], \nonumber\\ N&=&[M^2\a^2(x^2-1)^2+\a^2
(\a^2-M^2)(x^2-y^2)^2-(q^2+b^2)(1-y^2)^2]^2 \nonumber\\
&&-16\a^4\delta^2x^2y^2(x^2-1)(1-y^2), \nonumber\\
D&=&\{M^2\a^2(x^4-1)+\a^2(\a^2-M^2)(x^2-y^2)^2
+(q^2+b^2)(1-y^4)+2M\a x \nonumber\\
&&\times[\a^2(x^2-y^2)-M^2(1-y^2)]\}^2
+4\a^2\delta^2[\a(x^2-2x^2y^2+y^2)+Mx(1-y^2)]^2, \nonumber\\
W&=&M\a^2[(\a^2-M^2)(x^2-y^2)(3x^2+y^2) +M^2(3x^4+6x^2-1)+8M\a
x^3] \nonumber\\ &&-(q^2+b^2)[4\a xy^2-M(1-y^2)^2], \nonumber\\
\delta&=&\sqrt{M^2(\a^2-M^2)-q^2-b^2}, \quad \a=\frac{1}{2}R,
\label{extreme} \eea
where the prolate spheroidal coordinates ($x$, $y$) are related to
the cylindrical coordinates ($\rho$,~$z$) by the formulas
\be x=\frac{1}{2\a}(r_++r_-), \quad y=\frac{1}{2\a}(r_+-r_-),
\quad r_\pm=\sqrt{\rho^2+(z\pm\a)^2}, \label{xy} \ee
and where we have also given the explicit extremal form of the
function $I$ defining the magnetic potential $A_\varphi$ via
formula (\ref{tph}).

As it follows from (\ref{mm}), the magnetic field in our dihole
solution differs considerably from that of the particular $b=0$
specialization of the EMR solution considered in \cite{CLL}: in
the former case it has a dipole character, while in the latter
case it behaves itself like a magnetic octupole ($B_3=2q\delta$).
In Figs.~2 and 3 this difference is illustrated by the plots of
magnetic lines of force for two characteristic particular cases.

We end this section by observing that the singularity structure of
the solution (\ref{EFn}), which is determined by zeros of the
denominator $A+B$ of the Ernst potentials $\E$ and $\Phi$, is
essentially the same as that of the solution considered by
Cabrera-Munguia et al, i.e., no ring singularities appear for
$\rho>0$ in the positive mass case $M>0$ (this has been checked
numerically for a wide range of parameters of our solution), while
for $M<0$ there arise two massless ring singularities outside the
location of the sources, in agreement with a recent study
\cite{MRu} of the single KN spacetime endowed with negative mass.

\section{Conclusions}

In our paper we succeeded in elaborating a physically consistent
4-parameter model for stationary diholes whose exceptionality
consists in the fact that it is the {\it only} 4-parameter member
of the 5-parameter EMR class not containing magnetic charges
(whereas, on the other side, there is an infinite number of
4-parameter specializations of the latter class with individual
magnetic charges of the constituents). This model, generalizing
the known dihole electrostatic solution earlier obtained by
Emparan and Teo, is comprised of two identical (up to the sign of
charges) counter-rotating KN black holes supported from falling
onto each other by a massless strut. Curiously, its finding and
correct mathematical description has turned out to be a more
sophisticated task than in the case of counter-rotating equally
charged KN black holes represented by the BM solution because the
knowledge of a more general 5-parameter EMR solution was needed
for getting rid of the specific individual magnetic charges
initially present in the dihole components. The solution's
physical representation was advantageous for a direct check that
the binary configuration it describes really saturates the
Gabach-Clement inequality for interacting black holes.

Since the aforementioned inequality also takes into account the
possibility for the black holes to carry magnetic charges as
independent parameters, we would like to mention that our dihole
solution can be very easily generalized to the case when the two
KN constituents, besides the opposite electric charges, would have
arbitrary opposite magnetic charges too, thus representing a pair
of dyons \cite{Sch}. To introduce an arbitrary magnetic charge
${\mathcal B}$ into our dihole model, one only needs to make the
following substitutions in the formulas (\ref{sig})-(\ref{mfn}):
change $Q$ to ${\mathcal Q}=Q-i{\mathcal B}$, and $Q^2$ to
$|{\mathcal Q}|^2=Q^2+{\mathcal B}^2$ in all the occurrences. For
instance, our expression (\ref{sig}) for $\s$ will then assume the
form
\be \s=\sqrt{M^2-\left(\frac{M^2a^2[(R+2M)^2+4|{\mathcal Q}|^2]}
{[M(R+2M)+|{\mathcal Q}|^2]^2}+|{\mathcal
Q}|^2\right)\frac{R-2M}{R+2M}}. \label{sigB} \ee
We underline that the parameter ${\mathcal B}$ thus introduced
will be a genuine individual magnetic charge of the upper black
hole, and this can be readily verified by means of the formula
\be {\mathcal B}=\frac{1}{4\pi}\int_{H} \omega A_{t,z} d\varphi d
z. \label{mag} \ee

It is easy to see that the dyonic dihole model, which is of course
equivalent to the 5-parameter EMR solution, will also saturate the
Gabach-Clement inequality because the electric and magnetic
charges $Q_i$ and ${\mathcal B}_i$ enter that inequality only in
the combination $Q_i^2+{\mathcal B}_i^2$ \cite{Gab}. Mention also
that the introduction of the magnetic charge does not actually
modify seriously the Smarr mass formula (\ref{Sma}), provided the
substitutions described above are carried out properly in the term
$Q\Phi^H$.

\section*{Acknowledgments}

The authors would like to thank the referees for valuable comments
and suggestions. V.S.M. is grateful to the Relativity group of
School of Mathematical Sciences, Queen Mary, University of London,
with special thanks to Juan Valiente-Kroon, for their kind
hospitality and for providing excellent working conditions during
his visit, when part of this work was done. The research was
partially supported by CONACyT of Mexico.

\newpage

\begin{figure}
  \centering
    \includegraphics[width=12cm]{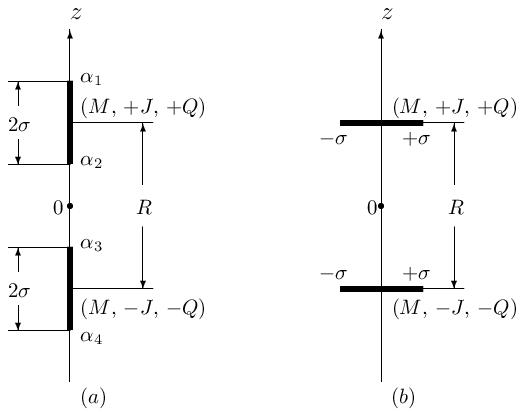}
  \caption{Location
of sources on the symmetry axis for two branches of the dihole
solution: (a) a black dihole configuration composed of two KN
black holes ($\s\ge0$); (b) a hyperextreme dihole configuration
composed of two superextreme KN constituents (pure imaginary
$\s$).}
  \label{fig1}
\end{figure}

\begin{figure}
  \centering
    \includegraphics[width=100mm]{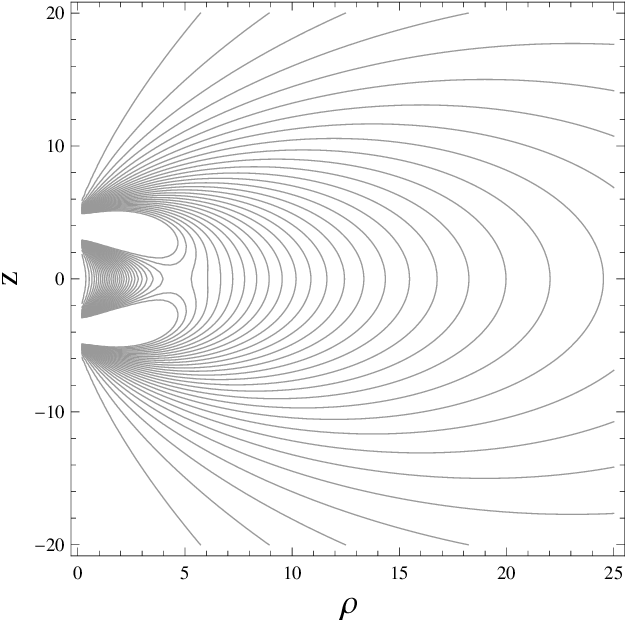}
  \caption{Magnetic
lines of force plotted for the dihole solution in the particular
case $R=8$, $M=3/2$, $a=1/8$, $Q=1/2$.}
  \label{fig2}
\end{figure}

\begin{figure}
  \centering
    \includegraphics[width=100mm]{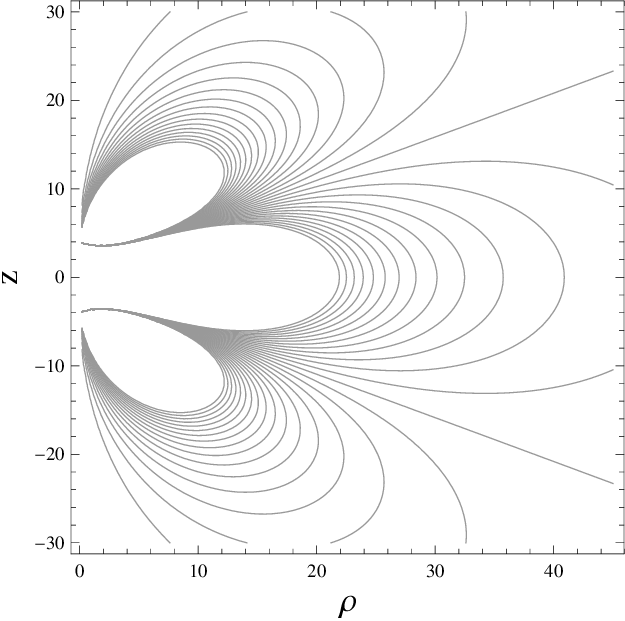}
  \caption{Magnetic
lines of force plotted for the specific EMR solution considered in
\cite{CLL}; the parameter choice (in the original notation for the
charge parameter) is $R=8$, $M=3/2$, $\s=1/4$, $Q_0=1/2$.}
  \label{fig3}
\end{figure}


\begin{references}
\bibitem{ETe} R. Emparan and E. Teo, \J{\NP}{B610}{190}{2001}.
\bibitem{Rei} H.~Reissner, \J{\APL}{355}{106}{1916}.
\bibitem{Nor} G.~Nordstr\"om, Proc. K. Ned. Akad. Wet. {\bf 20}, 1238 (1918).
\bibitem{CLL} I. Cabrera-Munguia, C. L\'ammerzahl, L. A. L\'opez and A. Mac\'ias,
\J{\PRD}{88}{084062}{2013}.
\bibitem{NCC} E. T. Newman, E. Couch, K. Chinnapared, A. Exton, A. Prakash, R.
Torrence, \J{\JMP}{6}{918}{1965}.
\bibitem{BMa} N. Bret\'on and V. S. Manko, \J{\CQG}{12}{1969}{1995}.
\bibitem{MRR} V.~S.~Manko, R. I. Rabad\'an and E. Ruiz, \J{\CQG}{30}{145005}{2013}.
\bibitem{Sma} L. Smarr, \J{\PRL}{30}{71}{1973}.
\bibitem{EMR1} F. J. Ernst, V.~S.~Manko and E. Ruiz, \J{\CQG}{24}{2193}{2007}.
\bibitem{SRo} J. Sod-Hoffs and E. D. Rodchenko, \J{\CQG}{24}{4617}{2007}.
\bibitem{Kom} A.~Komar, \J{\PR}{113}{934}{1959}.
\bibitem{Gab} M. E. Gabach Clement, \J{\CQG}{29}{165008}{2012}.
\bibitem{RMM} E. Ruiz, V.~S.~Manko and J. Mart\'in, \J{\PRD}{51}{4192}{1995}.
\bibitem{EMR2} F. J. Ernst, V.~S.~Manko and E. Ruiz, \J{\CQG}{23}{4945}{2006}.
\bibitem{Ern} F.~J.~Ernst, \J{\PR}{168}{1415}{1968}.
\bibitem{Sib} N. R. Sibgatullin, Oscillations and Waves in Strong
Gravitational and Electromagnetic Fields (Berlin: Springer, 1991).
\bibitem{MRRS} V.~S.~Manko, E. D. Rodchenko, E. Ruiz and B. I. Sadovnikov,
\J{\PRD}{78}{124014}{2008}.
\bibitem{BSi} R. Beig and W. Simon, \J{\PRSLA}{376}{333}{1981}.
\bibitem{SBe} W. Simon and R. Beig, \J{\JMP}{24}{1163}{1983}.
\bibitem{Sim} W. Simon, \J{\JMP}{25}{1035}{1984}.
\bibitem{HPe} C. Hoenselaers and Z. Perj\'es, \J{\CQG}{7}{1819}{1990}.
\bibitem{SAp} T. P. Sotiriou and T. A. Apostolatos, \J{\CQG}{21}{5727}{2004}.
\bibitem{Tom} A. Tomimatsu, \J{\PTP}{72}{73}{1984}.
\bibitem{MMR} V.~S.~Manko, J. Mart\'in and E. Ruiz, \J{\CQG}{23}{4473}{2006}.
\bibitem{Car} B. Carter, in: General Relativity, an Einstein Centenary Survey,
Eds. S. W. Hawking and W. Israel (Cambridge University Press,
Cambridge, 1979), p. 294.
\bibitem{CGM} J. A. C\'azares, H. Garc\'ia-Compe\'an and V. S. Manko, \J{\PLB}{662}{213}{2008}.
\bibitem{Var} G. G. Varzugin, \J{\TMP}{116}{1024}{1998}.
\bibitem{Isr} W. Israel, \J{\PRD}{15}{935}{1977}.
\bibitem{CHR} M. S. Costa, C. A. R. Herdeiro and C. Rebelo, \J{\PRD}{79}{123508}{2009}.
\bibitem{MRu} V.~S.~Manko and E. Ruiz, \J{\PTEP}{2013}{103E01}{2013}.
\bibitem{Sch} J. S. Schwinger, Science {\bf 165}, 757 (1969).

\end{references}
\end{document}